\documentclass[12pt,american]{article}
\usepackage[T1]{fontenc}
\usepackage[latin9]{inputenc}
\usepackage{geometry}
\usepackage{babel}
\usepackage{verbatim}
\usepackage{textcomp}
\usepackage{amsmath}
\usepackage{amssymb}
\usepackage{graphicx}
\usepackage[unicode=true,pdfusetitle,
 bookmarks=true,bookmarksnumbered=false,bookmarksopen=false,
 breaklinks=false,pdfborder={0 0 1},backref=false,colorlinks=false]
 {hyperref}

\makeatletter

\newcommand{\lyxmathsym}[1]{\ifmmode\begingroup\def\b@ld{bold}
  \text{\ifx\math@version\b@ld\bfseries\fi#1}\endgroup\else#1\fi}

\@ifundefined{date}{}{\date{}}
\usepackage{soul}
\usepackage{textgreek}

\makeatother

\begin{document}
\title{A detailed field theory for RNA-like molecules with periodic base
sequence}
\author{R. Dengler \thanks{ORCID: 0000-0001-6706-8550}}
\maketitle
\begin{abstract}
This work examines a field theory for RNA-like molecules in a good
solvent. The field theory is based on a lattice model for single-
and double-strand RNA with a periodic base sequence, and otherwise
contains all known relevant details (polymer types, polymer lengths
and interactions). As for the somewhat less explicit $O\left(n\right)$-symmetric
model there is a close relation to the conventional one-component
branched polymer and the associated Lee-Yang problem. We further elucidate
this relation. We derive exact results in the limiting cases of nearly
complete denaturation and nearly complete renaturation. The single-strand
critical exponent $\nu_{\varphi}$ is calculated in two-loop order.
\end{abstract}
The theory of RNA in a good solvent started with the classical work
of de Gennes \cite{Gennes1968}, who also introduced the approximation
of using a periodic base sequence, and emphasized the importance of
hairpin diagrams at low temperature \cite{Gennes1968}, which form
a kind of condensate.

The fundamental phenomenology is well known. In aqueous solution at
temperatures exceeding approximately $100\lyxmathsym{\textdegree}\mathrm{C}$,
single-strands predominate, while at lower temperatures double strands
become prevalent (renaturation). This transition occurs continuously,
lacking a critical temperature. 

This general picture has been consistently validated with many different
models, with various results for the statistics of polymer conformations.
We only mention a selection of recent studies \cite{Mueller2007,Rosa2019,S.Cocco2022,Vaupoti2023}.
The primary objective is to establish scaling laws and determine parameters
such as scaling exponents and radius of gyration.

The complexity of the issue is compounded by the nondeterministic
nature of base sequences and the necessity for relatively lengthy
polymers to achieve scaling limits.

It is clear however, that long polymers enforce long correlations
lengths, which normally entail critical phenomena. It thus should
be no surprise to find connections to the universality class of the
conventional branched polymer, at least in the case of a periodic
base sequence. 

\section*{\textquotedblleft All linear polymers are alike\textquotedblright}

Long linear polymers with self and mutual repulsion and other interactions
contain linear segments devoid of interactions, which also are long
if the interaction is weak. The probability distribution of the end
to end vector of such segments is a gaussian function. A complete
description of long linear polymers is nothing else than Gaussian
functions connected with interaction vertices, and one could say ``all
linear polymers are alike''.

A down-to-earth way to integrate over polymer configurations with
self- and mutual repulsion and other interactions is the path integral
of Edwards \cite{Edwards1965}, where the variables directly are the
positions of the monomers. A more convenient formulation of the problem
are more abstract (``second quantized'') field theories. Such field
theories may contain a lot of details, as a bookkeeping for polymer
types, interactions and lengths. This bookkeeping would be difficult
with the path integral of Edwards. But it is useful to keep in mind,
that in the end everything boils down to gaussian integrals of segments
in Feynman diagrams, which topologically are identical with polymer
conformations. This remark also encompasses branched ``linear''
polymers, for which a field theory has been derived and examined by
Lubensky and Isaacson in 1979 \cite{Lubensky1979}. 

As we have recently shown \cite{Dengler2022,Dengler2023} it is not
difficult to derive field theories for RNA-like branched polymers
with single and double strands from a lattice model. The RNA field
theory comes in two variants, a $O\left(n\right)$-symmetric variant
which requires to set $n=0$ in the end, and a variant with length
variables instead of $O\left(n\right)$ indexes without the $n=0$
limit. The field theory with length variables contains more information,
but is otherwise equivalent to the $O\left(n\right)$-symmetric model.

There is a close relation between the conventional branched polymer
of Lubensky and Isaacson and RNA with a periodic base sequence, and
such RNA appears to be a physical realization of the conventional
branched polymer universality class.

In \cite{Dengler2022,Dengler2023} we have examined the relation between
the models and the conventional branched polymer, and examined the
$O\left(n\right)$-symmetric model in more detail. This work fills
the last gap and examines the model with length variables. This leads
to results which cannot be derived or are difficult to derive from
the $O\left(n\right)$-model.

\section*{The model for single and double strand RNA}

The physical system are RNA like single- and double-strand polymers
with a periodic base sequence like GCGC... in a good solvent. The
case of a random base sequence is much more complicated \cite{Dengler2023}.
Here we examine in detail the most explicit field theory for such
polymer networks. The action integral reads \cite{Dengler2023}

\begin{align}
S_{\chi\varphi} & =\int_{x,s}\tilde{\varphi}_{s}\left(r+\partial_{s}-\nabla^{2}\right)\varphi_{s}+\tfrac{1}{2}\int_{x,s,s'}\chi_{s',s}\left(\tau+\partial_{s}-\partial_{s'}-\nabla^{2}\right)\chi_{s,s'}+\tfrac{\alpha}{2}\int_{x}\left(\int_{s}\chi_{s,s}\right)^{2}\label{eq:ActionPhiChiS}\\
 & \:-2g\int_{x,s,s'}\chi_{s,s'}\tilde{\varphi}_{s}\varphi_{s'}-\lambda\int_{x,s,s',s''}\chi_{s,s'}\chi_{s',s''}\chi_{s'',s}-H\int_{x,s}\chi_{s,s}.\nonumber 
\end{align}
Expressions like $\int_{x,s}...=\int\mathrm{d}^{d}x\int_{-\infty}^{\infty}\mathrm{d}s...$
are abbreviations for integrals over space $x$ and length variables
$s$, $s'$ and $s''$. To unclutter expressions length variables
are written as suffixes, space variables are omitted. The excluded
volume interactions $\left(\tilde{\varphi}\varphi\right)^{2},$ $\chi^{4}$
and $\tilde{\varphi}\varphi\chi^{2}$ are irrelevant at the upper
critical dimension $8$ and not written down. We also have ignored
``follower'' $\chi^{3}$ interactions \cite{kaviraj2022}, which
are influenced by the coupling constants $\lambda$ and $g$, but
do not act back on the ``leaders'' $\lambda$ and $g$.

Action $S_{\chi\varphi}$ has a simple and clear interpretation, and
a formal derivation from a lattice model is nothing special and not
really necessary. We mention once again that the topology of polymer
conformations and Feynman diagrams always are identical.

\begin{figure}

\centering{}\includegraphics[scale=3]{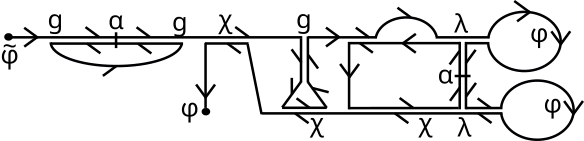}\caption{\label{fig:OnePhiUniverse}An RNA conformation in a universe generated
by a single-strand source ($\tilde{\varphi}$) and sink ($\varphi$).
There is a double-strand polymer $\chi$ with oppositely aligned single
strands. There are interactions with coupling constants $g$ and $\lambda$.
The two-point interaction with coupling constant $\alpha$ plays special
role. When length variables are Fourier transformed to \textquotedblleft frequencies\textquotedblright ,
then these frequencies run unchanged through the diagram, and effectively
only one frequency is involved.}
\end{figure}

But there are several unconventional aspects. First of all, the pair
of fields $\varphi$ and $\tilde{\varphi}$ describes the single-strand
polymer, the field $\chi$ describes the double strand polymer. Accordingly,
the fields $\varphi$ and $\tilde{\varphi}$ have one length variable,
the field $\chi$ has two length variables. Single-strand RNA molecules
have an internal direction ($5'$ to 3') and therefore there are two
fields, a $\varphi$ incoming into vertices and a $\tilde{\varphi}$
outgoing from vertices. Only oppositely aligned single RNA strands
form double strand RNA molecules $\chi$. The $s$ and $s'$ in $\chi_{s,s'}$
thus increment in different directions, and $\chi$ does not really
have a direction, see fig.(\ref{fig:OnePhiUniverse}). 

The question then is, which length variable is the first one in $\chi_{s,s'}$?
It is crucial for a local field theory to have a unique \emph{local}
convention. The rule is: in a $\chi_{s,s'}$ connected to a vertex
the first length variable $s$ is incoming, the second length variable
$s'$ is outgoing. This explains the transposition of the length variables
in the $\chi_{s',s}\chi_{s,s'}$ vertex in the action integral. The
$s'$ incoming on one side is outgoing on the other side, and vice
versa for $s$. This is similar in the interaction with coupling constant
$g$, which describes the transformation of two single-strands to
a double strand and the reverse effect. The incoming $s$ of $\chi_{s,s'}$
becomes the outgoing $s$ of $\tilde{\varphi}_{s}$, and vice versa
for $s'$.

The remaining interaction $-\lambda\chi^{3}$ describes the direct
transition of a double-strand polymer to two double-strand polymers,
see fig.(\ref{fig:OnePhiUniverse}). This interaction is generated
from the interaction with coupling constant $g$, and is marginal
at the upper critical dimension $d_{c}=8$. Note that $\lambda$ and
$g$ occur with positive sign in the statistical weight $e^{-S_{\chi\varphi}}.$

A special role is played by the strongly relevant harmonic term proportional
to $\alpha$. This term is generated by other interactions ($\lambda$,
$g$ or excluded volume), and must be taken into account from the
outset \cite{Lubensky1979,Dengler2022,Dengler2023,Parisi1981}. 

\subsection*{Frequency space}

The action (\ref{eq:ActionPhiChiS}) is local in space but non-local
in length variables $s$, $s'$ and $s''$. This means that interactions
between polymers are the same for any combination of length variables.
This translational invariance suggests to use the Fourier-transformed
representation. To length variables there correspond variables conjugate
under Fourier transformation which we call ``frequencies''. The
convention is $f_{s}=\int_{\omega}e^{-i\omega s}f_{\omega}$, with
$\int_{\omega}...=\int\mathrm{d}\omega/\left(2\pi\right)..$. In frequency
representation action (\ref{eq:ActionPhiChiS}) reads

\begin{align}
S_{\chi\varphi} & =\tfrac{1}{2}\int_{x\omega\omega'}\chi_{-\omega',-\omega}\left(\tau_{0}-i\left(\omega-\omega'\right)-\nabla^{2}\right)\chi_{\omega,\omega'}-\lambda\int_{x\omega\omega'\omega''}\chi_{\omega,-\omega'}\chi_{\omega',-\omega''}\chi_{\omega'',-\omega}\label{eq:ActionPhiChiOmega}\\
 & \qquad+\tfrac{\alpha}{2}\int_{x}\left(\int_{\omega}\chi_{\omega,-\omega}\right)^{2}+\int_{x\omega}\tilde{\varphi}_{-\omega}\left(r_{0}-\nabla^{2}-i\omega\right)\varphi_{\omega}-2g\int_{x\omega\omega'}\tilde{\varphi}_{\omega}\varphi_{\omega'}\chi_{-\omega,-\omega'}-H\int_{x,\omega}\chi_{\omega,-\omega}.\nonumber 
\end{align}
The translational invariance of the interactions implies that a frequency
along a single or double strand polymer runs unchanged through diagrams.
There remain no frequency integrals, except possibly at the very end
in a transformation back to length space and in closed frequency loops,
which drop out because length variables cannot increase consistently
in a loop \footnote{This is evident in length representation on a lattice, where each
propagator has a minimal length. In frequency representation loops
generate factors like $\delta\left(\omega=0\right)$, but this is
due to the fact that the minimal length is not accounted for in the
underlying strictly local field theory.}. The propagators are

\begin{align}
\left\langle \tilde{\varphi}_{\omega'}^{\boldsymbol{p}}\varphi_{\omega}^{\boldsymbol{k}}\right\rangle  & =\frac{\delta\left(\omega+\omega'\right)}{r_{0}+k^{2}-i\omega}\left(2\pi\right)^{d+1}\delta^{d}\left(\boldsymbol{p}+\boldsymbol{k}\right),\label{eq:Prop_Chi_omega}\\
\left\langle \chi_{\nu,\nu'}^{\boldsymbol{p}}\chi_{\omega,\omega'}^{\boldsymbol{k}}\right\rangle  & =\left(\frac{\delta\left(\omega+\nu'\right)\delta\left(\nu+\omega'\right)}{v_{k,\omega-\omega'}}-\alpha\frac{\delta\left(\omega+\omega'\right)}{v_{k,\omega-\omega'}}\frac{\delta\left(\nu+\nu'\right)}{v_{k,\nu-\nu'}}\right)\left(2\pi\right)^{d+2}\delta^{d}\left(\boldsymbol{p}+\boldsymbol{k}\right),\nonumber 
\end{align}
where $v_{k,\omega}=\tau_{0}+k^{2}-i\omega$. Action (\ref{eq:ActionPhiChiOmega})
and the propagators (\ref{eq:Prop_Chi_omega}) are very similar to
the $O\left(n\right)$-symmetric expressions of the model with $O\left(n\right)$
indexes instead of frequencies \cite{Dengler2022,Dengler2023,Parisi1981}.
Frequencies run unchanged through diagrams like $O\left(n\right)$
indexes, but there is nothing like a $n\rightarrow0$ limit.

\subsection*{Double polymer twisting}

Real RNA double strands are twisted in a rather rigid way. This implies
a stiffness against rotations around the longitudinal axis, which
is ignored in action (\ref{eq:ActionPhiChiS}). Such a type of stiffness
also is present in other linear polymers, and it is plausible that
this mechanical effect becomes less important for long polymers.

In the RNA system, however, there remains the topological effect caused
by an interchange of the two strands. A double polymer has two length
variables, one of them increments and the other decrements in a given
direction. These variables can be traced through any configuration
of interacting RNA. Whether a double strand is twisted or not can
make a difference for where a variable turns up in external lines.
This effect in a way is a nuisance, and it plays no role in renormalization
group calculations, which can be done with $\omega=0$ or, equivalently,
within the $O\left(n\right)$-symmetric model without length variables.
A correct field theory nevertheless must take the effect into account,
and we show how this is realized in action (\ref{eq:ActionPhiChiS}).

First of all, double polymer twisting does \emph{not} change the propagator.
By convention, in a $\chi_{s,s'}$ connected to a vertex, the first
length variable $s$ is incoming, the second length variable $s'$
is outgoing. The $\left\langle \chi\chi\right\rangle $ propagator
thus is agnostic to any twisting (it would be difficult anyway in
$d>2$ to say whether a segment is twisted or not). The twisting gets
generated by the interaction. In contrast to the usual case the interaction
\[
-\lambda\int_{x,\omega,\omega',\omega''}\chi_{\omega,-\omega'}\chi_{\omega',-\omega''}\chi_{\omega'',-\omega}
\]
is invariant only under cyclic permutations of the field factors.
Perturbation theory, however, allows contractions with any field.
As a result the propagation of length variables through an interaction
vertex generates alternatives.

It is also clear that any valid diagram can be generated in this way.
For a graph with directions, frequencies and twists, simply write
down the vertices with all details (fields with frequencies). This
are factors like $\chi_{\omega,-\nu}\chi_{\nu,-\rho}\chi_{\rho,-\omega}$.
Field-contractions then reproduce the graph. 

\subsection*{Debye structure function}

Scattering experiments measure the wavevector-dependent density-density
correlation function $D\left(k\right)$, normalized with the partition
sum $Z$. It is trivial but instructive to calculate $D\left(k\right)$
for a non-interacting (gaussian) linear polymer, for instance the
$\varphi$-polymer \cite{Debye1947}. This is not immediately relevant
for the RNA system, but illustrates some technical points. The Fourier
transformed mass density operator for any length variable $s$ is

\[
\rho_{k,s}^{\left(\varphi\right)}=\int\mathrm{d}^{d}p\tilde{\varphi}_{-p,s}\varphi_{p+k,s}/\left(2\pi\right)^{d}.
\]
Of interest is the $\rho$-$\rho$ correlation function in a universe
with a source $\tilde{\varphi}$ with length variable $0$ at the
origin, and a sink $\varphi$ with length variable $\ell$ anywhere
\footnote{This is one of the rare cases where a calculation in length space
is simpler.}, 
\begin{align*}
D_{\varphi}\left(k,\ell\right) & =\tfrac{1}{Z}\int_{p,u,s}\left\langle \tilde{\varphi}_{p,0}\rho_{-k,u}^{\left(\varphi\right)}\rho_{k,s}^{\left(\varphi\right)}\varphi_{0,\ell}\right\rangle \\
 & =\tfrac{2}{Z}\int_{0}^{\ell}\mathrm{d}ue^{-r_{0}u}\int_{u}^{\ell}\mathrm{d}se^{-\left(s-u\right)\left(r_{0}+k^{2}\right)}e^{-\left(\ell-s\right)r_{0}}\\
 & =\tfrac{2}{Z}e^{-\ell r_{0}}k^{-4}\left(e^{-\ell k^{2}}-1+\ell k^{2}\right).
\end{align*}
Here we have used the $\tilde{\varphi}\varphi$-Propagator $g_{k}\left(s\right)=\theta\left(s\right)e^{-s\left(r_{0}+k^{2}\right)}$.
The normalization with partition sum $Z=\int_{k}\left\langle \tilde{\varphi}_{k,0}\varphi_{0,\ell}\right\rangle =e^{-\ell r_{0}}$
removes the dependence on $r_{0}$. This is typical for polymers with
given total length. The parameter $r_{0}$ acts as chemical potential
for the constant polymer mass, and thus has no physical effect. 

\section*{Known results}

The field theory for branched polymers of Lubensky and Isaacson \cite{Lubensky1979}
is rather complicated, the $\chi$-sector of eq.(\ref{eq:ActionPhiChiS})
or eq.(\ref{eq:ActionPhiChiOmega}) or the corresponding $O(n)$-symmetric
model \cite{Dengler2022} are alternatives. A very similar other simpler
version can be found in ref.(\cite{Parisi1981}). It is known since
1981 that the critical point of ``conventional'' branched polymers
can be mapped to the Lee-Yang critical point in two less dimensions
\cite{Parisi1981}. The upper critical dimension is $8$, the critical
exponents in $d=3$ and $d=4$ are known exactly. As already noted
by Lubensky and Isaacson ``This theory is very rich'' \cite{Lubensky1979}. 

The additional single-strand field $\varphi$ in action (\ref{eq:ActionPhiChiOmega})
does not spoil the relation to the Lee-Yang system. This means that
peculiarities like a condensate
\begin{align}
\left\langle \chi_{\omega,\omega'}\right\rangle  & =2\pi\delta\left(\omega+\omega'\right)Q\left(\tau_{0},\omega\right),\label{eq:ChiCondensate}\\
Q\left(\tau_{0},\omega\right) & =\mathrm{a_{0}}-a_{1}\left(\tau_{0}-\tau_{0c}-2i\omega\right)^{\beta_{H}}+...,\nonumber 
\end{align}
Fisher renormalization of the critical exponents $\nu_{\chi}\rightarrow\nu_{H}=\nu_{\chi}\beta_{H}$
and $\gamma_{\chi}\rightarrow\gamma_{H}=\gamma_{\chi}\beta_{H}$,
as well as a renormalization group calculation only imitate the Lee-Yang
problem \cite{Dengler2022,Dengler2023}. Action (\ref{eq:ActionPhiChiOmega}),
however, contains more information. 

At the stable nontrivial fixed point with exactly $g_{*}=3\lambda_{*}$
the equivalence encompasses \emph{both} fields, i.e. $\chi$ and $\varphi$
have the same anomalous dimension \cite{Dengler2023}. The condensate
$\left\langle \chi\right\rangle \neq0$ shifts the $\chi$ mass parameter
as $\tau_{0}\rightarrow\tau_{0}-6\lambda Q$, but also the $\varphi$
mass parameter according to $r_{0}\rightarrow r_{0}-2gQ$. This means
that the $\chi$-$\varphi$ symmetry even is valid for $\tau_{0}\neq\tau_{0c}$
in the region $r_{0}-r_{0c}<2ga_{1}\left(\tau_{0}-\tau_{0c}\right)^{\beta_{H}}$.
But there is a new critical exponent $\nu_{\varphi}$, and of course,
action (\ref{eq:ActionPhiChiOmega}) contains precise information
about single- and double-strand polymer lengths.

\section*{Polymer lengths}

The field theory (\ref{eq:ActionPhiChiOmega}) with length variables
allows to derive the relative amounts of $\varphi$- and $\chi$-polymer,
at least in two limits. Consider a situation with (originally) a single
single-strand polymer of given length $\ell$ as in fig.(\ref{fig:OnePhiUniverse}).
Such a universe is generated by a single-strand source and sink, 
\begin{align}
\hat{\Phi}_{\ell} & =\int_{\omega}e^{-i\omega\ell}\int_{x,\omega'}\tilde{\varphi}_{\omega'}\left(0\right)\varphi_{\omega}\left(x\right)\equiv\int_{\omega}e^{-i\omega\ell}\hat{\Phi}_{\omega}.\label{eq:OnePhi_Universe}
\end{align}
The polymer starts at the origin with length variable $0$ and terminates
anywhere with fixed length variable $\ell$. The partition sum is
$Z=\left\langle \hat{\Phi}_{\ell}\right\rangle $, and in naive perturbation
series the parameter $\omega$ occurs only in the combinations $r_{0}-i\omega$
and $\tau_{0}-2i\omega$. This allows to write the total length as

\begin{align*}
\ell & =\tfrac{1}{Z}\int_{\omega}e^{-i\omega\ell}\partial_{\left(i\omega\right)}\left\langle \hat{\Phi}_{\omega}\right\rangle =\tfrac{1}{Z}\int_{\omega}e^{-i\omega\ell}\left(-\partial_{r_{0}}-2\partial_{\tau_{0}}\right)\left\langle \hat{\Phi}_{\omega}\right\rangle =\tfrac{1}{Z}\int_{\omega}e^{-i\omega\ell}\left\langle \left(\hat{m}_{\varphi}+\hat{m}_{\chi}\right)\Phi_{\ell,\omega}\right\rangle =\bar{m}_{\varphi}+\bar{m}_{\chi},
\end{align*}
where

\begin{align}
\hat{m}_{\varphi} & =\int_{x,\omega}\tilde{\varphi}_{-\omega}\varphi_{\omega},\label{eq:MassOperators}\\
\hat{m}_{\chi} & =\int_{x,\omega,\omega'}\chi_{-\omega',-\omega}\chi_{\omega,\omega'}\nonumber 
\end{align}
are the polymer mass operators.

\subsection*{Initial stage of pairing}

In the initial stage of pairing the single-strand polymer $\varphi$
dominates. This occurs when the parameter $\tau_{0}$ is large, away
from the $g_{*}=3\lambda_{*}$ fixed point. A plausible ansatz for
the action integral is a subset of the monomials of eq.(\ref{eq:ActionPhiChiOmega}),
\begin{align}
S_{\varphi\chi} & =\int_{x,\omega}\tilde{\varphi}_{-\omega}\left(r_{0}-\nabla^{2}-i\omega\right)\varphi_{\omega}+\tfrac{\tau_{0}}{2}\int_{x\omega\omega'}\chi_{-\omega',-\omega}\chi_{\omega,\omega'}+\tfrac{\alpha}{2}\int_{x}\left(\int_{\omega}\chi_{\omega,-\omega}\right)^{2}\label{eq:ActionDenat_PhiChi}\\
 & \qquad-2g\int_{x\omega\omega'}\tilde{\varphi}_{\omega}\varphi_{\omega'}\chi_{-\omega,-\omega'}+u_{\varphi}\int_{x}\left(\int_{\omega}\tilde{\varphi}_{-\omega}\varphi_{\omega}\right)^{2}-H\int_{x\omega}\chi_{\omega,-\omega}.\nonumber 
\end{align}
Only the excluded volume interaction for the single strand polymer
$\varphi$ with coupling constant $u_{\varphi}$ has been added. And
indeed, the coupling constants $u_{\varphi}$, $\tau_{0}$, $\alpha$
and $g$ are marginal in $S_{\varphi\chi}$ at the new upper critical
dimension $d_{c}=4$. Derivative terms $\chi\nabla^{2}\chi$ and $\chi i\omega\chi$,
the $\chi^{3}$ interaction, and $\chi^{2}$$\chi^{2}$ as well as
$\chi^{2}$$\varphi^{2}$ excluded-volume interactions are strongly
irrelevant and have been omitted. 

The double-strand field $\chi$ occurs only harmonically in $S_{\varphi\chi}$
and can be integrated out from the weight $e^{-S_{\varphi\chi}}$.
This is simple in perturbation theory with the $\chi\chi$-propagator
with $v_{k,\omega}=\tau_{0}$ from eq.(\ref{eq:Prop_Chi_omega}).
A contribution to $r_{0}$ of order $O\left(\alpha H\right)$ contains
a length loop and drops out. There remains $S_{\varphi\chi}$ without
$\chi$, but with new bare coupling constants
\begin{align*}
r_{1} & =r_{0}-2gH/\tau_{0},\\
u_{1} & =u_{\varphi}+2g^{2}\left(\alpha/\tau_{0}^{2}-1/\tau_{0}\right).
\end{align*}
The remaining field theory is equivalent to the $\varphi^{4}$-model
of de Gennes, with additional information in $r_{1}$ and $u_{1}$,
and additional length variable for the fields $\varphi$ and $\tilde{\varphi}$.
Of interest is the partition sum of the situation depicted in fig.(\ref{fig:OnePhiUniverse}),
the exact $\left\langle \tilde{\varphi}\varphi\right\rangle $ propagator.
This quantity can be calculated in leading power-law approximation
without any further approximation,

\begin{align*}
Z\left(r_{0},\tau_{0}\right)=\left\langle \hat{\Phi}_{\ell}\right\rangle  & \sim\int_{-\infty}^{\infty}\mathrm{d}\omega e^{-i\ell\omega}\left(r_{1}-r_{1c}-i\omega\right)^{-\gamma_{\varphi}}=\mathrm{const}\times\ell^{\gamma_{\varphi}-1}e^{-\left(r_{1}-r_{1c}\right)\ell}.
\end{align*}
The parameters $r_{1}$ and $\omega$ occur only in the combination
$r_{1}-i\omega$, and there is only wavevector $0$ in play. This
enforces the power law in the single remaining scaling variable $r_{1}-r_{1c}-i\omega$,
the exponent $\gamma_{\varphi}$ is the exponent of the $\varphi^{4}$-
$O\left(n=0\right)$ model. The frequency integral has been performed
with

\begin{equation}
\int_{-\infty}^{\infty}\tfrac{\mathrm{d}\omega}{2\pi}e^{-i\omega\ell}\left(B-i\omega\right)^{-k}=\ell^{k-1}e^{-B\ell}/\Gamma\left(k\right).\label{eq:GR_Formula}
\end{equation}
The average $\varphi$-mass as expected is $\bar{m}_{\varphi}=-\partial\ln Z/\partial r_{0}\sim\ell$
(in leading order). The average double strand mass follows as
\begin{equation}
\bar{m}_{\chi}=-\frac{\partial\ln Z}{\partial\tau_{0}}=-\left(\frac{\partial u_{1}}{\partial\tau_{0}}\frac{\partial}{\partial u_{1}}+\frac{\partial r_{1}}{\partial\tau_{0}}\frac{\partial}{\partial r_{1}}\right)\ln Z\sim O\left(\ln\ell\right)+\left(2gH-2g^{2}\frac{\partial r_{1c}}{\partial u_{1}}\right)\ell/\tau_{0}^{2}+O\left(\ell/\tau_{0}^{3}\right).\label{eq:mChiFinal}
\end{equation}
We are only interested in the contribution proportional to $\ell$.
The power law $\bar{m}_{\chi}\sim\ell/\tau_{0}^{2}$ for any dimension
looks somewhat trivial. It does not depend on the $\varphi^{4}$ critical
exponents, and thus there might be a simpler derivation or explanation.
The constraint $\bar{m}_{\varphi}+\bar{m}_{\chi}=\ell$ is not exactly
met, but this cannot be expected with the leading order approximation
for $\bar{m}_{\varphi}$.

\subsection*{Final stage of pairing}

\begin{figure}
\centering{}\includegraphics[scale=0.6]{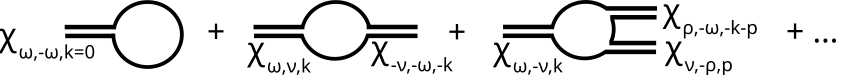}\caption{\label{fig:PhiElimination}Integrating out the single-strand field
$\varphi$ generates one-loop contributions to the $\chi$-sector
of the field theory.}
\end{figure}
In the final stage of pairing the double strand polymer prevails,
and it is useful to eliminate the single-strand field $\varphi$ from
action (\ref{eq:ActionPhiChiOmega}). This is possible because $\varphi$
occurs only harmonically, in particular in the three-point interaction
$S_{g}=-2g\int_{x\omega\omega'}\tilde{\varphi}_{\omega}\varphi_{\omega'}\chi_{-\omega,-\omega'}$.
The elimination generates a series of one-loop contributions, displayed
graphically in fig.(\ref{fig:PhiElimination}). These contributions
to the effective action for $\chi$ are of the type $S_{n,d}=-g^{n}I_{n,d}\left(r_{0},y\right)\int\chi^{n}$,
with $n\geq1$ fields $\chi$ and

\begin{align*}
I_{n,d}\left(r_{0},y\right) & =\mathrm{const}\int^{\Lambda}\frac{\mathrm{d}^{d}p}{\left(r_{0}+p^{2}+y\right)^{n}}\overset{d\leq2n}{\sim}\left(r_{0}+y\right)^{-\left(n-d/2\right)},\\
y & =O\left(-i\omega_{i},k_{i}^{2}\right).
\end{align*}
The integrals $I_{n,d}$ are IR convergent and analytic in $r_{0}+y$
for $d>2n$. Integrals $I_{n\geq4,d}$ are singular for dimensions
$d\leq8$, but $S_{n\geq4,d}$ is strongly irrelevant. The power law
for $I_{n,d}\left(r_{0},y\right)$ is valid for $\left|r_{0}+y\right|\ll\Lambda^{2}$,
that is when the single-strand polymer is not completely suppressed.
The new parameters of the remaining $\chi$-field-theory are 
\begin{align}
H_{1} & =H+c_{1}I_{1,d}\left(r_{0},0\right),\nonumber \\
\tau_{1} & =\tau_{0}-c_{2}I_{2,d}\left(r_{0},0\right),\label{eq:ChiParams_with_r0}\\
\lambda_{1} & =\lambda+c_{3}I_{3,d}\left(r_{0},0\right).\nonumber 
\end{align}
Contributions smaller by a factor $O\left(y/r_{0}\right)$ can be
neglected. The relation of the remaining effective double-strand field
theory to the Lee-Yang system remains valid, but the dependence of
the parameters (\ref{eq:ChiParams_with_r0}) on $r_{0}$ contains
information about the single-strand polymer. 

It recommends itself to now consider a universe with a $\chi$-source
and a $\chi$-sink, generated by

\begin{equation}
\hat{\Omega}_{\ell}=\int_{x,s}\chi_{0,0}\left(0\right)\chi_{\ell,s}\left(x\right)=\int_{\omega}e^{-i\omega\ell}\int_{x,\nu,\nu'}\chi_{\nu,\nu'}\left(0\right)\chi_{\omega,\omega'=0}\left(x\right).\label{eq:ChiUniverse}
\end{equation}
The source is at the origin and has length variables $0$, the sink
is anywhere but has first length variable $\ell$, the second length
variable is arbitrary (it automatically gets fixed to $-\ell$). The
partition sum

\begin{equation}
Z=\left\langle \hat{\Omega}_{\ell}\right\rangle \sim\int_{\omega}e^{-i\omega\ell}\left(\tau_{1}-\tau_{1c}-i\omega\right)^{-\gamma_{H}}\sim\ell^{\gamma_{H}-1}e^{-\left(\tau_{1}-\tau_{1c}\right)\ell}\label{eq:Z_WithoutPhi}
\end{equation}
is a $\chi$-propagator with wavevector zero. The parameters $\tau_{1}$
and $\omega$ occur only in the combination $\tau_{1}-i\omega$. This
enforces the power law in the single remaining scaling variable $\tau_{1}-\tau_{1c}-i\omega$,
the exponent $\gamma_{H}=\gamma_{\chi}\beta_{H}$ is the (Fisher-renormalized)
exponent of the Lee-Yang model. The integral in has been performed
with eq.(\ref{eq:GR_Formula}). With eq.(\ref{eq:MassOperators})
there follow the average masses

\begin{align*}
\bar{m}_{\chi} & =-2\partial\ln Z/\partial\tau_{0}=2\ell,\\
\bar{m}_{\varphi} & =-\frac{\partial\ln Z}{\partial r_{0}}\cong-\frac{\partial H_{1}}{\partial r_{0}}\frac{\ln Z}{\partial H_{1}}-\frac{\partial\tau_{1}}{\partial r_{0}}\frac{\partial\ln Z}{\partial\tau_{1}}-\frac{\partial\lambda_{1}}{\partial r_{0}}\frac{\partial\ln Z}{\partial\lambda_{1}}\sim c_{1}I_{2,d}\left(r_{0}\right)\frac{\ln Z}{\partial H_{1}}+2c_{2}I_{3,d}\left(r_{0},y\right)\ell+...
\end{align*}
It remains to calculate $X=\partial\ln Z/\partial H_{1}$. The partition
sum (\ref{eq:Z_WithoutPhi}) depends on $H_{1}$ via $\tau_{1c}$,
but it is better to use $X=\tfrac{1}{Z}\partial Z/\partial H_{1}=\tfrac{1}{Z}\left\langle \hat{\Omega}_{\ell}\int_{x,\omega}\chi_{\omega,-\omega}\right\rangle $,
see eq.(\ref{eq:ActionPhiChiOmega}) and eq.(\ref{eq:ChiUniverse}).
This are three $\chi$-$\chi$ propagators $\left(\tau_{1}-\tau_{1c}-i\omega\right)^{-\gamma_{H}}$
with a vertex function $\Gamma_{\chi\chi\chi}\sim\left(\tau_{1}-\tau_{1c}-i\omega\right)^{2\gamma_{H}-1}$
in between (see appendix for the exponent). Again there is only wavevector
$0$ in play, and the factors are powers of the single remaining scaling
variable $\tau_{1}-\tau_{1c}-i\omega$. With the help of formula (\ref{eq:GR_Formula})
one finds $X\sim\ell$ and 
\begin{equation}
\bar{m}_{\varphi}\sim c_{1}'I_{2,d}\left(r_{0}\right)\ell+2c_{2}I_{3,d}\left(r_{0},y\right)\ell+...\sim c_{1}''\ell r_{0}^{-\left(2-d/2\right)}+c_{2}''\ell r_{0}^{-\left(3-d/2\right)}+...\label{eq:mPhiFinal}
\end{equation}
The dependence of $\bar{m}_{\chi}$ on the bare coupling constant
$\lambda_{1}$ is still smaller. Universal quantities like $\gamma_{H}$
do not depend on $\lambda_{1}$, nonuniversal quantities like amplitudes
and temperature shifts like $\tau_{1c}$ are regular functions of
$\lambda_{1}$.

The power law $\bar{m}_{\varphi}\sim\ell/\sqrt{r_{0}}$ in $d=3$
for large $r_{0}$ looks somewhat trivial. It does not depend on the
Lee-Yang critical exponents, and there might be a simpler derivation.
The constraint $\bar{m}_{\varphi}+\bar{m}_{\chi}=2\ell$ is not exactly
met, but this cannot be expected with the leading power-law approximation
for $Z$ and $\bar{m}_{\chi}$. The calculation has the air of thermodynamics,
and there might be a simpler explanation for the result (\ref{eq:mPhiFinal}).

The contributions $S_{n,d}$ from $\varphi$-elimination to the effective
$\chi$-sector look as if this would give a field theory not equivalent
to the Lee-Yang problem for $r_{0}=r_{0c}$. However, this is an illusion.
The $\varphi$-loops cannot contain an $\alpha$-propagator (eq.(\ref{eq:Prop_Chi_omega})),
and thus are less singular and do not contribute to the critical point
\cite{Dengler2023}. 

\subsection*{Intermediate range}

Only one of the polymer masses is independent because of the constraint
$\bar{m}_{\varphi}+\bar{m}_{\chi}=\ell$, and one can consider the
ratio $\bar{m}_{\varphi}/\text{\ensuremath{\ell}}$ as a degree of
freedom. This ratio usually depends on many physical quantities: temperature,
acidity of solvent, concentrations of ions etc. The parameters $r_{0}$
and $\tau_{0}$ depend on these physical quantities, but always must
change in such a way that the constraint is fulfilled. If $\tau_{0}$
is chosen as independent parameter then $r_{0}=r_{0}\left(\tau_{0},\ell\right)$
is a kind of hyperbola in the $\tau_{0}$-$r_{0}$-plane. A large
$\tau_{0}$ diminishes $\bar{m}_{\chi}$, and this is compensated
by $r_{0}$ becoming small. If $\bar{m}_{\varphi}$ approaches the
maximal value $\ell$, then the influence of $r_{0}$ disappears ($i\omega\sim1/\ell$
dominates $r_{0}$), and the physical quantities only act via $\tau_{0}.$
This and the symmetric situation lead to the two limiting cases above,
with either large $r_{0}$ or large $\tau_{0}$ as only and \emph{free}
physical parameter.

It is more difficult to get quantitative results for the transitional
range. Starting from the dissociated limit (the $\varphi^{4}$ model),
double-strand mass $\bar{m}_{\chi}\sim\ell/\tau_{0}^{2}$ increases
rapidly. Starting from the paired limit, single-strand mass $\bar{m}_{\varphi}\sim\ell/\sqrt{r_{0}}$
increases much more slowly. 

Here the stable fixed point with $g_{*}=3\lambda_{*}$ comes into
play. At this fixed point $\chi$ and $\varphi$ together form a branched
polymer of the conventional type ($\chi$ and $\varphi$ have same
anomalous dimension \cite{Dengler2023}), with comparable amounts
of $\bar{m}_{\varphi}$ and $\bar{m}_{\chi}$. This at least indicates
that the Lee-Yang regime extends into the transitional regime.

The main contributions to the polymer masses near this stable fixed
point originate from a mass operator (eq.(\ref{eq:MassOperators}))
insertion into the condensate $\left\langle \chi\right\rangle $ (hairpin
diagrams). The calculation for $r_{0}$ small and $\tau_{0}$ not
too large is similar to eq.(\ref{eq:mPhiFinal}). There are three
$\chi$-$\chi$ propagators with a vertex $\Gamma_{\chi\chi\chi}$
in between, only the dependence on $r_{0}$ is missing. Both masses
are of order $O\left(\ell\right)$, with nonuniversal amplitudes.

\section*{Two-loop calculation}

The critical exponent $\nu_{\varphi}$ does not show up in the leading
approximation of polymer length $\bar{m}_{\varphi}$ in the final
stage of single-strand pairing (the calculation based on the $O\left(n\right)$
model of \cite{Dengler2023} is misleading). However, a more precise
value for $\nu_{\varphi}$ might still be of interest. 

An ab initio two-loop calculation for the Lee-Yang field theory is
demanding and not worth the effort. Higher-loop-calculations nowadays
are a science by itself, and there are elaborate tools for nearly
every step of the calculation. Our goal here is more modest. We show,
how known results can be combined to get the exponent $\nu_{\varphi}$
in two loop order. 

From the $5$-loop results \cite{Borinsky2021} one can take over
the functions $\beta\left(\lambda_{6}\right)$, $\eta_{\varphi}\left(\lambda_{6}\right)$,
where the suffix $6$ at the coupling constant denotes the upper critical
dimension. When modified minimal subtraction is used in both cases,
then the coupling constants of the Lee-Yang- and polymer-problem are
related by $\lambda_{6}^{2}=\left(3-\epsilon/2\right)\lambda_{8}^{2}$,
where $\epsilon=8-d$ \cite{AIM76}. This allows to deduce the functions
$\beta\left(\lambda_{8}\right)$ and $Z_{\varphi}\left(\lambda_{8}\right)$.

There remains the calculation of the $\Gamma_{\varphi\varphi}$ vertex
function with a $\varphi^{2}$ insertion. This does not lead to any
new two-loop diagrams (the diagrams are depicted as $E$, $a_{1},...,\text{\ensuremath{a_{5}} }$in
the appendix of \cite{Dengler2023}). An integration by parts algorithm
allowing to reduce the integrals to two simple master integrals is
concisely described in \cite{BierWein2003}. With the requisite algebra
implemented in the C++-Ginac framework we find 
\[
\nu_{\varphi}=\tfrac{1}{2}+\tfrac{\epsilon}{36}+\tfrac{11}{1458}\epsilon^{2}+...
\]
This still does not allow a plausible extrapolation to $d=3$, but
at least provides a check for a more sophisticated calculation.

\section*{Fields with unique scaling dimensions}

An peculiar aspect of action (\ref{eq:ActionPhiChiOmega}) and similar
actions for branched polymers is that the field $\chi$ does not have
a unique scaling dimension, c.f. $\left\langle \chi\chi\right\rangle \sim k^{-2}$
and $\left\langle \chi\chi\right\rangle \sim k^{-4}$ in eq.(\ref{eq:Prop_Chi_omega}).
And there are other, related, peculiarities. The parameter $\alpha$
does not have a finite fixed point. This is not really a problem,
because in the end $\alpha$ occurs only in the effective couplings
constants $\lambda\alpha^{1/2}$ and $g\alpha^{1/2}$. But a naive
dimensional analysis is not sufficient to find consistent canonical
dimensions of the fields and the upper critical dimension.

The root of the complications is the $\alpha$-term. The $\alpha$-term
is harmonic, but can occur only once in a propagator line, and thus
more behaves as an interaction than as a propagator. It is of interest
to transform action (\ref{eq:ActionPhiChiS}) or (\ref{eq:ActionPhiChiOmega})
to an equivalent form without these peculiarities.

The essential observation is that every loop in Feynman diagrams contributing
to the critical point contains exactly one $\alpha$-term \cite{Lubensky1979,Parisi1981}.
Cutting such a diagram at the $\alpha$-terms generates a tree diagram,
and the $\alpha$-term in this sense acts as the source of fluctuations,
connecting solutions of the classical equation of motion \cite{Parisi1981}. 

The first step is to separate the ``diagonal'' and off-diagonal
components of $\chi$,
\begin{equation}
\chi_{\omega\nu}=\hat{\chi}_{\omega\nu}+2\pi\delta\left(\omega+\nu\right)\rho_{\omega},\label{eq:ChiAsChiHatRho}
\end{equation}
with $\rho_{\omega}=\chi_{\omega,-\omega}$. \footnote{The notation $\rho_{\omega}$ is more compact, while $\chi_{\omega,-\omega}$
makes frequency conservation explicit.} Inserting eq.(\ref{eq:ChiAsChiHatRho}) into eq.(\ref{eq:ActionPhiChiOmega})
leads to expressions like $\delta\left(\omega=0\right)$ and $\delta\left(\omega\right)^{2}$,
which can be regularized to $\delta\left(0\right)=Na$ and $\delta\left(\omega\right)^{2}=Na\delta\left(\omega\right)$
by temporarily using discrete frequencies $\omega\in\left\{ 0,\tfrac{2\pi}{a},...,\tfrac{2\pi}{a}\left(N-1\right)\right\} $,
where $a$ is monomer length and $N$ a large number. The result is

\begin{align}
S' & =S_{\hat{\chi}\varphi}+S_{\hat{\chi}\rho},\label{eq:ActionRhoChi_0}\\
S_{\hat{\chi}\varphi} & =\tfrac{1}{2}\int_{x\omega\omega'}\hat{\chi}_{-\omega',-\omega}\left(\tau_{0}-i\left(\omega-\omega'\right)-\nabla^{2}\right)\hat{\chi}_{\omega,\omega'}+\int_{x,s}\tilde{\varphi}_{s}\left(r+\partial_{s}-\nabla^{2}\right)\varphi_{s}\nonumber \\
 & \qquad-\lambda\int_{x,\omega,\omega',\omega''}\hat{\chi}_{\omega,-\omega'}\hat{\chi}_{\omega',-\omega''}\hat{\chi}_{\omega'',-\omega}-2g\int_{x\omega\omega'}\tilde{\varphi}_{\omega}\varphi_{\omega'}\hat{\chi}_{-\omega,-\omega'},\nonumber \\
S_{\hat{\chi}\rho} & =V\left(\rho\right)-\tfrac{A}{2}\int_{x}\left(\int_{\omega}\rho_{\omega}\right)^{2},\nonumber \\
V\left(\rho\right) & =\tfrac{1}{2}\int_{x\omega\omega'}\rho_{\omega}\left(\tau_{0}-2i\omega-\nabla^{2}\right)\rho_{\omega}-\lambda'\int_{x\omega}\rho_{\omega}^{3}-3\lambda''\int_{x\omega\nu}\hat{\chi}_{\omega,-\nu}\hat{\chi}_{\nu,-\omega}\rho_{\omega}-2g\int_{x\omega}\tilde{\varphi}_{\omega}\varphi_{-\omega}\rho_{\omega}.\nonumber 
\end{align}
Here we also have written $A=-\alpha>0$, this sign change is to be
corrected at the end. The remaining transformation concerns $S_{\hat{\chi}\rho}$
and resembles the transformation in the case of conventional branched
polymers \cite{Parisi1981}, only the ``frequency'' variables are
new. With the help of an auxiliary field $h$ one can write
\[
e^{-S_{\hat{\chi}\rho}}=\int_{-\infty}^{\infty}\mathrm{D}he^{-\tfrac{1}{2A}\int_{x}h^{2}}e^{-V\left(\rho\right)+\int_{x}h\int_{\omega}\rho_{\omega}}.
\]
The equation of motion $\delta V/\delta\rho-h=0$ for $\rho$ can
now be enforced with a $\delta$-function, 

\[
e^{-S_{\hat{\chi}\rho}}=\int_{-\infty}^{\infty}\mathrm{D}he^{-\tfrac{1}{2A}\int_{x}h^{2}}\int_{-i\infty}^{i\infty}\mathrm{D}\tilde{\rho}e^{-\int_{x\omega}\tilde{\rho}_{\omega}\left(\delta V/\delta\rho-h\right)}\det\left(\delta V/\delta\rho\right).
\]
Performing the average over $h$ leads to
\[
S_{\hat{\chi}\rho}=-\tfrac{A}{2}\int_{x}\left(\int_{\omega}\tilde{\rho}_{\omega}\right)^{2}+\int_{x\omega}\tilde{\rho}_{\omega}\delta V/\delta\rho.
\]
In a last step the sign of $A$ must be corrected. This can be achieved
by making the coupling constants imaginary \cite{Parisi1981}. The
final result is

\begin{align}
S_{\hat{\chi}\rho}' & =\int_{x}\left\{ -\tfrac{A}{2}\left(\int_{\omega}\tilde{\rho}_{\omega}\right)^{2}+\int_{\omega}\tilde{\rho}_{\omega}\left(\tau_{0}-2i\omega-\nabla^{2}\right)\rho_{\omega}-3i\lambda'\int_{\omega}\tilde{\rho}_{\omega}\rho_{\omega}^{2}\right\} \label{eq:ActionRhoChiOmega}\\
 & \qquad-3i\lambda''\int_{x\omega\nu}\tilde{\rho}_{\omega}\hat{\chi}_{\omega,-\nu}\hat{\chi}_{\nu,-\omega}-2ig\int_{x\omega}\tilde{\varphi}_{\omega}\varphi_{-\omega}\tilde{\rho}_{\omega}.\nonumber 
\end{align}
Here we have ignored the functional determinant. Its only effect
would be to remove closed loops of length variables  \footnote{The determinant can be taken into account with the help of fermionic
fields.}.

In $S_{\hat{\chi}\varphi}+S_{\hat{\chi}\rho}'$ now everything is
as usual. According to a dimensional analysis $\alpha$ is dimensionless,
and one can set $A=1.$ The propagators are (c.f. eq. (\ref{eq:Prop_Chi_omega}))

\begin{align*}
\left\langle \tilde{\rho}_{\nu}^{k}\rho_{\omega}^{p}\right\rangle  & =\frac{\delta\left(\omega-\nu\right)}{v_{k,2\omega}}\left(2\pi\right)^{d+1}\delta^{d}\left(\boldsymbol{k}+\boldsymbol{p}\right),\\
\left\langle \rho_{\nu}^{k}\rho_{\omega}^{p}\right\rangle  & =\frac{1}{v_{k,2\omega}v_{p,2\nu}}\left(2\pi\right)^{d}\delta^{d}\left(\boldsymbol{k}+\boldsymbol{p}\right),\\
\left\langle \hat{\chi}_{\omega,\omega'}^{k}\hat{\chi}_{\nu,\nu'}^{p}\right\rangle  & =\frac{\delta\left(\omega+\nu'\right)\delta\left(\omega'+\nu\right)}{v_{k,\omega-\omega'}}\left(2\pi\right)^{d+2}\delta^{d}\left(\boldsymbol{k}+\boldsymbol{p}\right).
\end{align*}
The interaction $\lambda\hat{\chi}^{3}$ is irrelevant at the upper
critical dimension $8$, the interactions $\lambda'\tilde{\rho}\hat{\chi}\hat{\chi}$,
$\lambda''\tilde{\rho}\rho^{2}$and $g\tilde{\varphi}\varphi\tilde{\rho}$
are marginal. In a situation as in fig.(\ref{fig:OnePhiUniverse})
there is no $\hat{\chi}$ field, except possibly in frequency loops.
It follows that $\hat{\chi}$ can be ignored, in particular $\hat{\chi}$
does not contribute to the critical behavior. 

The fact that the fields $\rho$, $\tilde{\rho}$ and $\hat{\chi}$
do have unique scaling dimensions does not necessarily simplify calculations.
An alternative to action (\ref{eq:ActionRhoChiOmega}) without the
$\tilde{\rho}$ field is to simply ignore $\hat{\chi}$ in action
(\ref{eq:ActionRhoChi_0}). The merit of the formulation (\ref{eq:ActionRhoChiOmega})
is of conceptional nature.

\section*{Conclusions}

In this work we have closed the last gap and examined the RNA field
theory with length variables, which explicitly contains all known
relevant details. In all calculations a periodic base sequence was
assumed. 

We have shown that it is possible to derive exact results for the
polymer masses for the initial and final stage of single-strand pairing.
It would be of interest to compare the results (\ref{eq:mChiFinal})
and (\ref{eq:mPhiFinal}) with other calculations or experiments.

The gyration radius $\sqrt{\ell}$ of the underlying conventional
branched polymer also was derived in \cite{Vaupoti2023} by means
of Flory type arguments and found to agree with simulations. This
is a confirmation for our field theory, the $\chi$-sector of which
is closely related to the model introduced by Parisi \cite{Parisi1981}.

It would be of interest to measure the gyration radius of real RNA
with periodic base sequence near complete renaturation. A scaling
law $\sqrt{\ell}$ would indicate that this system belongs to the
conventional branched-polymer universality class. Finite size scaling
might be useful to extrapolate to long polymer chains.

It is questionable whether RNA with random base sequence has much
in common with real RNA. But incorporating randomness into the field
\cite{Dengler2023} theory might show some tendency in comparison
to the somewhat unnatural case of a periodic base sequence studied
here.\bigskip{}

\bibliographystyle{habbrv}
\bibliography{DoublePolymer_Synopsis}
\bigskip{}

\appendix

\section*{Appendix}

\addcontentsline{toc}{section}{Appendix}

\subsection*{Scaling dimensions}

In order to reduce confusion, we derive here the scaling law of the
$3$-point vertex function $\Gamma_{\chi\chi\chi}\sim\tau^{2\gamma_{H}-1}$
in $\omega$-representation, which was used in the estimation of the
polymer mass $\bar{m}_{\varphi}$ in the final stage of single-strand
pairing, eq.(\ref{eq:mPhiFinal}).

There are different critical exponents in play, there is the Lee-Yang
and the branched-polymer system, and there are the $O\left(n\right)$
field theories and the field theories with length- or frequency-variables.

In $\omega$-representation there are no $\omega$-integrals in vertex
functions, and the vertex functions have the same scaling dimensions
as their $O\left(n\right)$ counterparts. It thus is simplest to simply
ignore all $\omega$-arguments. The scaling dimensions of $\chi^{2}$
by definition is $\chi^{2}\sim k^{d-1/\nu_{\chi}}.$ Because of Fisher
renormalization $\tau\rightarrow\tau^{\beta_{H}}$ \cite{Lubensky1979}
wavevector and temperature variable are related by $k\sim\tau^{\nu_{H}}$
with $\nu_{H}=\nu_{\chi}\beta_{H}$ (instead of the usual $k\sim\tau^{\nu_{\chi}}$).

The $3$-point vertex function $\Gamma_{\chi\chi\chi}$ is a self
energy $\Gamma_{\chi\chi}$ with an additional $\lambda\chi^{3}$
interaction with a (truncated) external leg, which apart from a constant
factor is a $\Gamma_{\chi\chi}$ with an inserted $\int\mathrm{d}^{d}x\chi^{2}\sim k^{-1/\nu_{\chi}}$.
By definition $\Gamma_{\chi\chi}\sim\tau^{\gamma_{H}}$ and thus $\Gamma_{\chi\chi\chi}\sim\tau^{\gamma_{H}}k^{-1/\nu_{\chi}}\sim\tau^{2\gamma_{H}-1}$,
where $\beta_{H}+\gamma_{H}=1$ was used (there are many relations
between the critical exponents) \cite{LubMcKane1981}.

\pagebreak
\end{document}